\documentclass{emulateapj}

\usepackage{graphicx,xspace}

\newcommand{\chandra}{{\sl Chandra}\xspace}
\newcommand{\suzaku}{{\sl Suzaku}\xspace}

\newcommand{\integral}{{\sl INTEGRAL}\xspace}
\newcommand{\swift}{{\sl Swift}\xspace}

\begin{document}

\title{High-resolution X-ray Spectra Of The Symbiotic Star SS73 17}

\author{R.N.C. Eze\altaffilmark{1,2}, G.J.M. Luna\altaffilmark{1}, and R.K. Smith\altaffilmark{1}}
\altaffiltext{1}{Smithsonian Astrophysical Observatory, 60 Garden Street,
  Cambridge, MA 02138}
\email{reze@head.cfa.harvard.edu, romanus.eze@unn.edu.ng}
\altaffiltext{2}{University of Nigeria, Nsukka, Enugu State, Nigeria}

\begin{abstract}
SS73 17 was an innocuous Mira-type symbiotic star until \integral and \swift discovered its bright hard X-ray emission, adding it to the small class of "hard X-ray emitting symbiotics." \suzaku observations in 2006 then showed it emits three bright iron lines as well, with little to no emission in the 0.3--2.0 keV bandpass. We present here followup observations with the \chandra HETG and \suzaku that confirm the earlier detection of strong emission lines of Fe K$\alpha$ fluorescence, \ion{Fe}{25} and \ion{Fe}{26} but also show significantly more soft X-ray emission. The high resolution spectrum also shows emission lines of other highly ionized ions as \ion{Si}{14} and possibly \ion{S}{16}.  In addition, a reanalysis of the 2006 \suzaku data using the latest calibration shows that the hard (15-50 keV) X-ray emission is brighter than previously thought and remains constant in both the 2006 and 2008 data.  

The G ratio calculated from the \ion{Fe}{25} lines shows that these lines are thermal, not photoionized, in origin.  With the exception of the hard X-ray emission, the spectra from both epochs can be fit using thermal radiation assuming a differential emission measure based on a cooling flow model combined with a full and partial absorber.  We show that acceptable fits can be obtained for all the data in the 1-10 keV band varying  only the partial absorber.  Based on the temperature and accretion rate, the thermal emission appears to be arising from the boundary layer between the accreting white dwarf and the accretion disk. 
\end{abstract}


\section{Introduction}
\label{sec:intro}

Symbiotic stars are interacting binaries whose components are a red giant and hot companion which accretes mass from the stellar wind of the red giant, producing a blue continuum that ionizes the surrounding gas. In most symbiotic stars the accretor is a white dwarf (WD), although some symbiotic stars ({\it e.g.} GX 1+4) have a neutron star companion \citep{CR97}. In X-ray wavelengths, symbiotic stars were detected as moderately bright sources in the ROSAT/All Sky Survey. \citet{Murset97} examined 16 symbiotic stars seen with ROSAT and categorized them into three classes: (1) super soft emission from the photosphere of the white dwarf ($\alpha$-type), (2) emission from an optically-thin thermal ($kT \sim 0.2$ keV) plasma possibly due to colliding winds from the two stars or to accretion ($\beta$-type), and (3) an ill-defined category of relatively hard X-ray sources ($\gamma$-type).  The origin of this hard emission remains uncertain to this day, and although not in the \citet{Murset97} survey, the ROSAT observation of SS73 17 suggests it would likely have been placed in category (3), albeit with some uncertainty due to the large column density (N$_{\rm H} = 1.8 \times 10^{22} $\,cm$^{2}$) required to fit ROSAT data \citep{Smith08}.

Earlier in 2005, both \integral \citep[IGRJ10109-5746,][]{Revnivtsev06a} and \swift \citep[SwiftJ101103.3-574814,][]{Atel669} independently discovered a hard X-ray source which \citet{Atel715} quickly identified with CD-57~3057 (SS73~17).  Combined with a small group of objects -- CH Cyg \citep{Ezuka98}, RT Cru \citep{Atel528, Luna07}, and T CrB \citep{Atel669, Luna08} -- these sources form the ``hard X-ray emitting symbiotics''. Afterwards, a dedicated \suzaku observation of SS73 17 revealed the presence of strong iron lines in the $6-7$\,keV region \citep{Smith08} which pointed to a thermal origin for the X-ray emission. Another remarkable feature in the X-ray spectrum of SS73~17 is its highly absorbed soft X-ray emission (with N$_{\rm H} > 10^{23}$\,cm$^{-2}$\,; \citet{Smith08}).

The origin of the hard and weak soft X-ray emission have remained a mystery which we hoped to address with a combination of \chandra HETG and \suzaku observations. In  \S\ref{sec:obs}, we summarize our observations and data processing, \S\ref{sec:results} contains our spectral analysis, while \S\ref{sec:timing} presents the timing analysis.  A discussion of our results is found in \S\ref{sec:disc}.

\section{Observation and Data Analysis}
\label{sec:obs}

We observed SS73~17 with the \chandra High Energy Transmission Grating Spectrometer (HETGS) and the \suzaku X-ray Imaging Spectrometer (XIS) and Hard X-ray Detector (HXD). The HETG data have a spectral resolution of 0.012 \AA\ \& 0.023 \AA \,FWHM for the High and Medium Energy Gratings (HEG, MEG), respectively.  The \suzaku XIS data cover the $0.3 - 12$\,keV range with $\sim 150$\,eV resolution, while the HXD extends the energy coverage from 10 to 600$\,$keV, although the source could not be detected above 50 keV.   Details on the \suzaku instrumentation and calibration can be found in the \suzaku website \footnote{http://heasarc.gsfc.nasa.gov/docs/suzaku/astroegof.html}.
Our goal was to obtain simultaneous \suzaku and \chandra observations in both soft and hard X-rays.  Due to the differing pointing constraints of the two telescopes, the observations were close together but not overlapping in time (see Table~\ref{tab:obs}). 

\begin{deluxetable}{llcccc}
\tablecolumns{5}
\tablewidth{0pc}
\tablecaption{SS73 17 Observations}
\tablehead{ \colhead{Satellite} & \colhead{ObsId} & \colhead{Start Date} &\colhead{Start Time [UT]} & \colhead{Exposure Time [ks]} } 
\startdata
\suzaku & 401055010 & 06/05/2006 & 05:13:12 & 17.9  \\
\chandra & 8967 & 10/23/2008 & 9:08:24 & 34.6 \\
\chandra & 10765 & 11/5/2008 & 9:43:58 & 19.3 \\
\suzaku & 403043010 & 11/11/2008 & 16:30:00 & 19.5 \\
\chandra & 10859 & 1/20/2009 & 6:48:27 & 10.3 \\
\chandra & 10793 & 1/21/2009 & 16:34:23 & 16.7 \\
\chandra & 10860 & 2/24/2009 & 18:24:54 & 13.0 \\
\chandra & 10869 & 2/28/2009 & 3:41:25 & 6.4 \\
\enddata
\label{tab:obs}
\end{deluxetable}

The \chandra and \suzaku observations dates and exposure times are shown in Table~\ref{tab:obs}. 
Multiple (six in total) \chandra observations were done for operational reasons, but this fortunately allowed us to study the long-term variability of the source.  CIAO 4.1 and CALDB 4.1 were used for the \chandra data analysis.  Each of the six observations was independently analyzed to obtain the Response Matrix Functions (RMFs) using the \texttt{mkgrmf} script and the Ancillary Response Matrices (ARFs) files using the
\texttt{fullgarf} script. After confirming that the spectral shape remains constant in all six observations to within our ability to measure it, we merged data from the six observations to increase the signal-to-noise ratio and obtain more reliable spectral fits.
We used \texttt{add\_grating\_orders} and
\texttt{add\_grating\_spectra} scripts to merge the grating orders and
grating spectra respectively.  We expect to lose some sensitivity to line broadening due to this merging, which is acceptable since our results do not hinge upon the line width. The merged data has a total exposure time of 100.023 ks with 3,557 and 3,729 counts in the HEG and MEG arms respectively between $0.3-10$\,keV.

Both \suzaku observations were analyzed using version 2 of the standard \suzaku pipeline software; in the case of ObsID 401055010, this is an update from the version used in \citet{Smith08}. In both cases, the pointing direction was chosen to center SS73~17 on the HXD detector which has the effect of reducing the effective area of the XIS by 10\% due to vignetting. We extracted all events within 4$^{\prime}$ of the source for the XIS detectors to produce our source spectra. Response matrices were generated for the XIS detectors using version 2009-02-28 of the \texttt{xisrmfgen} and effective area files for HXD-nominal pointing using \texttt{xisarfgen}. The XIS background spectra were extracted from a circular region with no apparent sources that was offset from both the source and the corner calibration sources.

We used only the HXD/PIN detector because the source was not bright enough to be detected in the HXD/GSO. We used the PIN response matrix appropriate for our data as generated by \suzaku team. We obtained the PIN background events from HXD/PIN Background files for V2.x Processed Data
\footnote{http://heasarc.gsfc.nasa.gov/docs/suzaku/analysis/pinbgd.html}
and then used \texttt{mgtime} to merge the GTIs to get common values for the background and source event files. \citet{Smith08} filtered out a significant number of PIN counts due to calibration uncertainties in part of the data; this has now been addressed and therefore our re-analysis includes these counts.  Dead time in the HXD/PIN was corrected using the \texttt{hxdtcor}\ routine. The cosmic X-ray background was included as an additional model term in our spectral modeling, based on the all-sky fits of \citet{Boldt87}.

Light curves were constructed by extracting photons from a circular region centered on the source with 240$^{\prime\prime}$ radius in the case of \suzaku data and 10$^{\prime\prime}$ in the \chandra zeroth order images. In the \suzaku observations we extracted background light curves from photons in three circles of 140$^{\prime\prime}$ radius around the source region. An annulus from 15$^{\prime\prime}$ to 30$^{\prime\prime}$ around the central source was used to construct background light curves in the \chandra observations. 

\section{Spectral Analysis and Results}
\label{sec:results}

\subsection{ \chandra HETG Spectral Analysis and Results}

\begin{figure}
\begin{center}
\includegraphics[totalheight=2.5in]{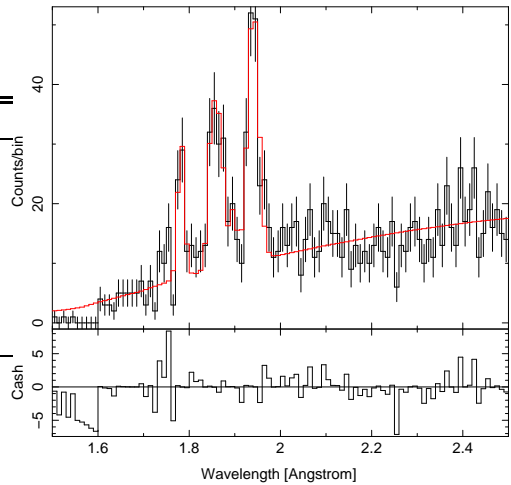}
\end{center}
\caption{[Top] HETG spectrum of SS73~17 in the $1.5 - 2.5$\,\AA\ region, showing the fit to the iron lines using a bremsstrahlung continuum and six Gaussian lines. Note that the \ion{Fe}{25}\ triplet in the middle is heavily blended, unlike the Fe K$\alpha$ fluorescence line. [Bottom] Cash statistic residuals for the fit. (See the electronic edition of the Journal for a color version of this and other figures).\label{fig:FeLines}}
\end{figure}

\begin{figure}
\begin{center}
\includegraphics[totalheight=2.5in]{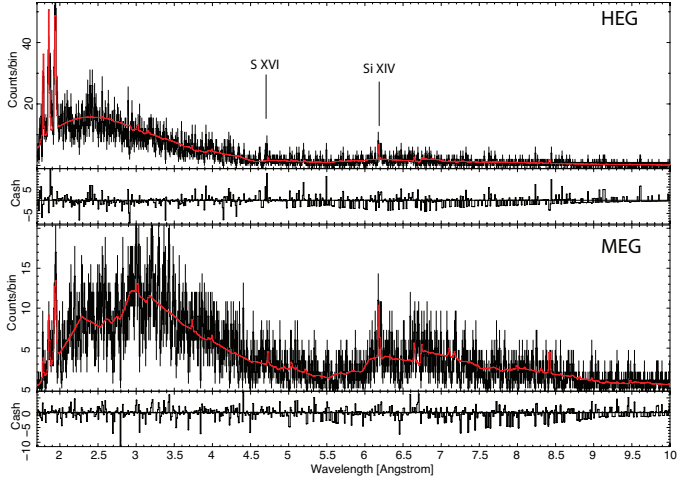}
\end{center}
\caption{[Top] HEG spectrum of SS73~17 in the $1.5 - 10$\,\AA\ region, showing
the fit to a two-temperature APEC thermal plasma with both a partial covering and full absorption model. [Bottom] Same, for MEG data.  \label{fig3}}
\end{figure}

Spectral analysis of the \chandra\ HETG data was performed using the Interactive Spectral Interpretation System  (ISIS)\footnote{http://space.mit.edu/CXC/ISIS/}.  In Figures~\ref{fig:FeLines} and \ref{fig3} we show the grating X-ray spectra of SS73~17 from the \chandra\ HETG observation. As shown in Figure~\ref{fig:FeLines}, we modeled the spectrum using a thermal bremsstrahlung continuum with six Gaussian lines in the narrow $\lambda = 1.5-2.5$\,\AA\ bandpass.  The best-fit values for the observed iron line positions, strengths, and Gaussian widths  ($\sigma$) are given in Table~\ref{tab:FeKlines}; all errors in this and other tables are 90\% confidence limits. 

\begin{table}
\caption{\chandra HETG line fit Parameters for $\lambda = 1.5 - 2.5$\AA \label{tab:FeKlines}}
\begin{tabular}{llll}
\hline \hline Line ID$^{*}$ & Line Position  & Observed Flux & Width \\
      & \AA & $10^{-5}$ ph/cm$^2$/s & m\AA \\ \hline 
Fe K & 1.939$\pm$0.003 & 6.5(-1.2, +2.2) & 10.70$\pm$ 3 \\
Fe XXV(w) & 1.849$\pm$0.005 & 3.2(-1.3, +1.6) & $<$ 13.6 \\
Fe XXV(x+y) & 1.868$\pm$0.005 & 2.7(-1.1, +3.0) & $<$ 13.6 \\
Fe XXV(z) & 1.896$\pm$0.005 & 1.0(-0.7, +1.8) & $<$ 13.6 \\
Fe XXVI &  $1.783\pm0.006$ & $3.3\pm1$           & $<$ 11.7 \\ \hline
\end{tabular}

\noindent $^{*}$ Fe XXV line labels follow the nomenclature of \citet{Gabriel72} \\
\end{table}

The relative positions of the \ion{Fe}{25}\ triplet lines were fixed, although their overall position was
allowed to vary. This is necessary because the individual lines are not resolved in the HEG. As a result of this partial blending, the individual line flux errors have large uncertainties although the overall detection of the \ion{Fe}{25} lines is strong.  We also fixed the relative positions of the \ion{Fe}{26} doublet; the mean position is given in Table~\ref{tab:FeKlines}. 

We detected a few additional emission lines from \ion{Si}{14} and possibly \ion{S}{16} in the soft X-ray spectrum.  These lines could not have been detected in the original \suzaku results due to the larger absorption seen during that observation.  However, the single-temperature ($\sim 9.3$\,keV) thermal plasma model used in \citet{Smith08} would not generate these lines for any value of the absorption column density.  \ion{Si}{14}, for example, has its peak emission at 1.36 keV, and at 9.3 keV these lines are only 5\% of their peak value \citep{Smith01}. We therefore fit the overall spectrum ($1.5-10$\,\AA) using an absorbed (full and partial cover absorption) two-temperature (using the APEC \citep{Smith01} thermal plasma code) model. Our best-fit model shows that the system is heavily absorbed (N$_{\rm H}=14.65\times 10^{22}$cm$^{-2}$ for the partial covering absorber and N$_{\rm H}=1.59\times 10^{22}$cm$^{-2}$\ for the full covering absorber) and the two plasmas have well-differentiated temperatures (with $kT_{1}=9.90$\,keV and $kT_{2}=1.12$\,keV) (see Table~\ref{tab:ChandraFit} and Figure~\ref{fig:FeLines}).

\begin{table}
\caption{\chandra HETG line fit Parameters for $\lambda = 1.5 - 10$ \AA\ (APEC thermal)\label{tab:ChandraFit}}
\begin{tabular}{lll}
\hline \hline Parameter & Unit & Value \\    \hline
N$_{\rm H}$ full & 10$^{22}$cm$^{2}$ & 1.59(-0.05, +0.16) \\
N$_{\rm H}$ partial &10$^{22}$cm$^{2}$ & 14.65(-1.56, +1.77) \\
Covering fraction & & 0.84$\pm$ 0.01 \\
kT$_1$ & keV & 9.90(-0.35, +1.20) \\
kT$_2$ & keV  & 1.12(-0.42, +0.16) \\ 
Abundance$_1$ & solar & 0.64(-0.10, +0.11) \\
Abundance$_2$ & solar & 1.00$\pm$0.02 \\ \hline
\end{tabular}
\end{table}

\begin{table}
\caption{\chandra HETG line fit Parameters for additional lines\label{tab:OtherLines}}
\begin{tabular}{llll}
\hline \hline Line ID & Line Position  & Observed Flux & Width \\
      & \AA & $10^{-5}$ ph/cm$^2$/s & m\AA \\ \hline 
S XVI & 4.71 (-0.05,+0.02) & $< 2.5$ & fixed at 0 \\
Si XIV & 6.19 $\pm$0.01 & $0.29\pm0.1$ & $22\pm10$ \\ \hline
%Mg XII & 8.448 $\pm$0.006 & 1.28(-1.0, +0.9) & \\ \hline
\end{tabular}
\end{table}

\subsection{ \suzaku Spectral Analysis and Results}

\begin{figure}
\includegraphics[totalheight=2.2in]{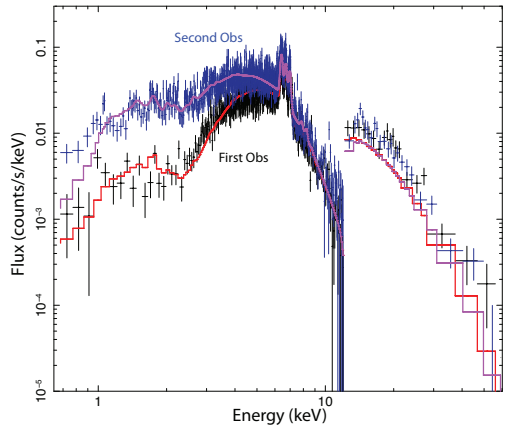}
\caption{Joint fit of the two \suzaku spectra fit with a {\tt mkcflow} model.  The fit is somewhat poor both at low and high energies, although it provides a good description between 2.0--8.0 keV. \label{fig:Joint_first}}
\end{figure}

The \suzaku data covers a much broader energy range than the HETG, albeit at a lower resolution. As it turns out, fitting this broad range with a single physically-motivated model was quite difficult. As we will show here, we were unable to find a convincing fit over the entire bandpass, although we considered a number of possibilities. We were motivated in part by comparisons of the \suzaku\ spectra from the first and second epoch which showed that the spectra in the 6--50 keV bandpass, including the iron lines, were unchanged while the lower-energy spectra changed dramatically; see Figure~\ref{fig:Joint_first}.  This suggested that the underlying X-ray source was relatively constant, while the absorber changed.  We note that the soft X-ray flux in the second \suzaku observation (F$_{\rm X}$[0.5-2 keV] $= 1.7\times10^{-13}$\,erg cm$^{-2}$s$^{-1}$) is in much better agreement with the original ROSAT observation ($\sim 4\times10^{-13}$\,erg cm$^{-2}$s$^{-1}$\ based on a pointed count rate of 0.0253 cts/s) than the first \suzaku observation \citep{Bickert96, Smith08}.  

The simple two-temperature model described above did an adequate job in the 0.5--10 keV bandpass, but did not fit the 10--50 HXD PIN data.  The indications from the HETG two-temperature fits of both a hard ($\sim 10$\,keV) and soft ($\sim 1$\,keV) thermal plasma, combined with the nature of the symbiotic system suggested a better model would include a range of temperatures due to accretion onto the white dwarf.  Following earlier papers on symbiotic stars \citep{Luna07, Stute09}, we used the {\tt mkcflow}\ model \citep{mkcflow} to fit the spectrum.  This model, originally developed for cooling-flow galaxy clusters, assumes a cooling plasma whose differential emission measure per unit temperature is inversely proportional to the bolometric luminosity, thus cooling equal amounts at each temperature.  The fit parameters include the peak temperature in the plasma and the total cooling (or accretion) flux per unit time.  The low temperature was held fixed at 0.0808 keV after our fits showed the results were insensitive to the exact value used.  

\begin{table}
\caption{\suzaku Joint$^{*}$\ Fit Parameters\label{tab:SuzakuFit}}
\begin{tabular}{llll}
\hline \hline Parameter & Unit & Obs 1 & Obs 2 \\ \hline 
N$_{\rm H}$ partial & 10$^{22}$cm$^{2}$ & 18$\pm$0.4 & 9.5$\pm$0.3 \\
Covering fraction & & 0.97$\pm$0.02 & 0.90$\pm$0.01 \\
N$_{\rm H}$ full      &10$^{22}$cm$^{2}$ & 0.92$\pm$0.06 & joint \\
$T_{high} $     & keV & 37$\pm$2 & joint \\
$\dot{M}^{**}$ & $10^{-10} M_{\sun}$/yr & $1.5\pm0.2^{+4.5}_{-1.1}$ & joint \\ \hline
\end{tabular}

\noindent $^{*}$\,Joint parameters were forced to be the same for both fits. \\
\noindent $^{**}$\,Errors are statistical plus those due to distance uncertainty.
\end{table}

The best-fit model gives an accretion rate $\dot{M}=1.5^{+4.5}_{-1.1}\times10^{-10}M_\sun$/yr; the errors are dominated by the uncertainties in the distance, taken here to be $500^{+500}_{-250}$ pc as estimated by \citet{Smith08}.  Although this range is of the same order of magnitude as typical accretion rates for symbiotics \citep{Livio84}, this model cannot be considered a complete success, as both the soft and hard X-rays are relatively poorly fit -- in particular, the model dramatically underestimates the hard ($>10$ keV) X-ray flux (although we note it is a better fit than the two-temperature model).  

\begin{figure}
\includegraphics[totalheight=2.7in]{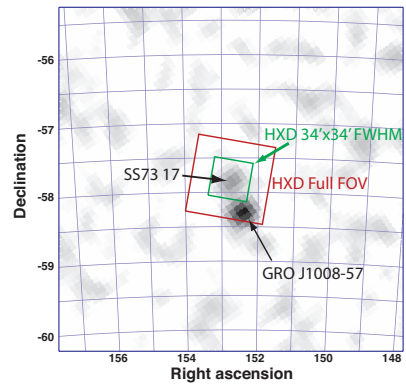}
\caption{14--25 keV BAT survey data around SS73 17 showing the \suzaku HXD/PIN field of view (both FWHM and the extreme edges of the collimator).  Although GRO J1008-57 is within the formal limits of the HXD/PIN detector, the detection efficiency of this far off-axis source is $<10$\%, and the apparent brightness of the source is due to its transient outbursts, none of which occurred during the \suzaku observation.
\label{fig:BATsurvey} }
\end{figure}

The failure of the model to fit at low energies is difficult to diagnose due to the low count rates and the large number of possible sources -- this could be due to photoionization of the M giant wind from the compact object, a colliding wind shock, a more complex source model than the {\tt mkcflow}\ model, or some combination of these.  Without more data in the soft band, it will be difficult to untangle this.  However, the failure of the model at higher energies is both more dramatic and has a smaller number of possible causes.  One immediate possibility is that there is a contaminating source in the field of view of the HXD PIN besides the cosmic X-ray background.  As noted by \citet{Atel669}, there is a transient HMXB, GRO J1008-57, $31^{\prime}$\ from SS73 17.  Fortunately, the \swift\ BAT survey is now available via {\sl Skyview}\footnote{http://skyview.gsfc.nasa.gov}, and we were able to use this along with the BAT lightcurves of selected transient sources \footnote{http://swift.gsfc.nasa.gov/docs/swift/results/transients/} to determine the impact of GRO J1008-57.  As shown in Figure~\ref{fig:BATsurvey}, this source is outside the FWHM field of view of the PIN, although it is still within the range where emission can leak into the source at the 10-15\% level (see Figure 8.3 of the \suzaku proposer's guide\footnote{http://heasarc.gsfc.nasa.gov/docs/astroe/prop\_tools/suzaku\_td/node11.html}).  The apparent brightness of the GRO J1008-57 relative to SS73 17 is due to the former's transient nature; during both \suzaku observations the source was in quiescence.  An analysis of the lightcurves of both sources using the BAT transient website shows that the average 15-50 keV flux of SS73 17 is $\sim 1.7\pm0.2\times10^{-11}$\,erg cm$^{-2}$s$^{-1}$, while the average quiescent GRO J1008-57 flux is $(1.9\pm0.2)\times10^{-11}$\,erg cm$^{-2}$s$^{-1}$.  A simple power-law fit to the HXD PIN data alone from both \suzaku observations returns a 15-50 keV flux (not including particle or cosmic X-ray backgrounds) of $(2.9\pm0.4)\times10^{-11}$\,erg cm$^{-2}$s$^{-1}$.  This difference from the $Swift$/BAT result could be due to a fluctuation in the source itself, or due to calibration differences between the BAT and the PIN; we note there is an 18\% offset between the \suzaku XIS and PIN detectors due to calibration uncertainties\footnote{See \suzaku Memo 2008-06 available at ftp://legacy.gsfc.nasa.gov/suzaku/doc/xrt/suzakumemo-2008-06.pdf}.

As contamination seemed unlikely to resolve the question of the missing flux, we considered two alternative sources of hard X-ray emission: (1) a reflection component from the accretion disk, or (2) an additional highly-absorbed power-law component from a possible unseen jet in the system.  

Adding a reflection component resulted in a good fit from 1-50 keV, but unfortunately required an unphysical solid angle for reflection of $R \equiv \Omega/2\pi > 3$, even for a face-on disk.  \citet{Reeves09}, who faced similar circumstances when modeling the quasar PDS 456, found a plausible fit to the spectrum of the Seyfert 2 with $R = 1.3$.   However, while a value slightly in excess of 1 for the parameter R can be understood if the underlying source is in fact more absorbed than the fit suggests, but a value $R > 3$\ would require a significant increase in the underlying accretion rate.  \citet{PR85}, in the context of accreting cataclysmic variable systems, found that accretion rates much above $2\times10^{-10}$\,M$\sun$/yr would quench the hard X-ray emission because the region around the boundary layer of the accretion disk would become optically thick.  
We also found that the addition of a power-law component with $\Gamma = 1.04\pm0.13$\ and intrinsic F$_{\rm X}$(15-50 keV) $= 2.1\times10^{-11}$\,erg cm$^{-2}$s$^{-1}$, absorbed by the same two components as the cooling flow plasma, also significantly improves the fit in the 15-50 keV bandpass, although with only the weak physical justification that there could be an (unseen) jet in system generating a relatively flat synchrotron spectrum.

\section{Timing Analysis}
\label{sec:timing}

\begin{figure*}
\includegraphics[totalheight=2.2in]{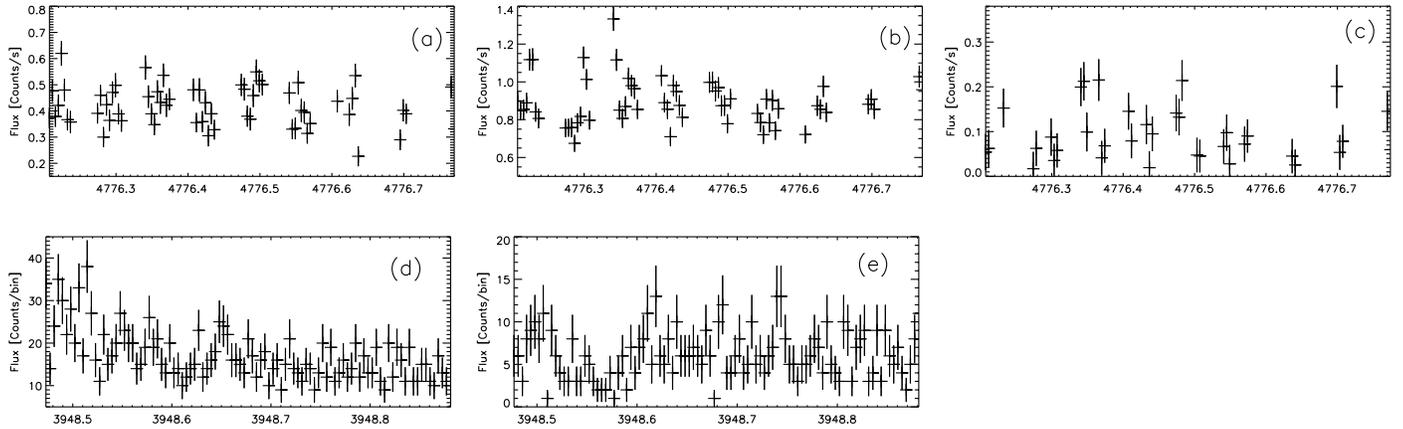}
\caption{SS73 17 X-ray light curves. The first row shows the light curves extracted from a \suzaku observation (ObsID 403043010, see Table~\protect{\ref{tab:obs}}\ for details) and binned at 360 s in three energy bands, 0.3--5.0, 5.0--10.0 and 12.0--50.0 keV in panels (a), (b) and (c) respectively. The second row shows \chandra zeroth order (ObsID 8967, see Table~\protect{\ref{tab:obs}}\  for details) light curves also binned at 360 s in two energy bands, 0.3--2.0 and 2.0--10.0 keV in panels (d) and (e) respectively. As mentioned in the text, the source shows strong variability on time scales of minutes, as expected from viscous instabilities in the internal region of an accretion disk.\label{fig:lc}}
\end{figure*}

The cornerstone of our timing analysis is the comparison between the ratio of measured fractional rms variation, $s$, to that expected from Poisson fluctuations alone, $s_{exp}$ \citep[see][]{Luna07}. In Figure \ref{fig:lc} we show the light curves from \suzaku and \chandra, respectively. All the light curves were extracted in three energy bands -- 0.3--5.0 , 5.0--10.0 and 12.0--50.0 keV -- dividing the spectral regions where absorption is important and binned at 360 s and 64 s. The values of the $s/s_{exp}$ ratio are listed in Table \ref{tab:timing}. 

The value of $s/s_{exp}$\ indicates that SS73 17 has strong stochastic variability in all the energy ranges observed. During both \suzaku observations the fractional amplitude of the stochastic variations was higher in the 5.0--10.0 keV band than in the 0.3--5.0 and 12.0--50.0 keV ranges, both in the 64 s (except for ObsID 403043010) and 360 s binned light curves. \chandra\ observations show that SS73 17 is sometimes more variable in the high energies than in the soft band.
%, e.g during the first pointing (ObsID 8967), the $s/s_{exp}$\ was higher in the soft X-rays. 

We also searched for modulated emission in the light curves. Using the arrival times of the source events we calculated the Z$^{2}_{1}$\ (Rayleigh) statistic \citep{Buccheri83} for frequencies $f_{min} = 1/T_{exp}$\ to $f_{max} = 1/2t_{frame}$\ with $\Delta f = 1/T_{exp}$, where $T_{exp}$\ is the exposure time as listed in Table~\ref{tab:obs}, $t_{frame}$\ is the read out time (2.54204 s in the case of \chandra\ and 8 s in the case of \suzaku\ XIS). We do not detect periods in the light curves. Using Eq. 1 from \citet{Luna07}, during \chandra\ observations we were sensitive to oscillations with a fractional amplitudes from 13 \% to 26 \%. In this case, we analyze the six observation segments separated. The higher number of counts obtained with \suzaku\ allows us to search for pulsations with a fractional amplitude as low as 4\%, although again we did not find modulation in the light curves.

\begin{table*}
\caption{Value of the $s/s_{exp}$ for each observation listed in Table 1\label{tab:timing}}
% \begin{center}
\begin{tabular}{lccccccc}
\hline \hline  & \multicolumn{3}{c}{64 s} &  & \multicolumn{3}{c}{360 s} \\
  & 0.3 - 5.0 keV & 5.0 - 10 keV & 12.0 - 50.0 keV &   & 0.3 - 5.0 keV & 5.0 - 10 keV & 12 - 50 keV \\  \hline
\hbox to 1.0in{\suzaku\leaders\hbox to 0.5em{\hss.\hss}\hfill} &  & & &  \\
\hbox to 1.0in{401055010\leaders\hbox to 0.5em{\hss.\hss}\hfill} & 3.58&4.44&1.18& &3.80 &8.24 &1.67  \\
\hbox to 1.0in{403043010\leaders\hbox to 0.5em{\hss.\hss}\hfill} & 2.17 &1.90&1.22& &3.30 &4.82 &1.94  \\
\hbox to 1.0in{\chandra\leaders\hbox to 0.5em{\hss.\hss}\hfill} &  & & &  \\
\hbox to 1.0in{8967\leaders\hbox to 0.5em{\hss.\hss}\hfill} &1.26  &1.23 &$\cdots$& &1.83&1.61&$\cdots$  \\
\hbox to 1.0in{10765\leaders\hbox to 0.5em{\hss.\hss}\hfill} &1.25  &1.50 &$\cdots$& &1.21 &1.88&$\cdots$  \\
\hbox to 1.0in{10859\leaders\hbox to 0.5em{\hss.\hss}\hfill} &1.10  &1.12 &$\cdots$& &1.57&1.96&$\cdots$  \\
\hbox to 1.0in{10793\leaders\hbox to 0.5em{\hss.\hss}\hfill} &1.25  &1.12 &$\cdots$& &1.29 &1.15&$\cdots$  \\
\hbox to 1.0in{10860\leaders\hbox to 0.5em{\hss.\hss}\hfill} &1.17  &1.15 &$\cdots$& &1.02 &1.47&$ \cdots$  \\
\hbox to 1.0in{10869\leaders\hbox to 0.5em{\hss.\hss}\hfill} &1.40  &1.78 &$\cdots$& &2.35 &2.28&$\cdots$  \\
\hline
\end{tabular}
% \end{center}
\end{table*}
% \vspace{1in}

\section{Discussion and Conclusions}
\label{sec:disc}

Symbiotic stars have been poorly studied because in the optical only limited information can be obtained; most of the interesting activity in these systems is, in fact, occurring in other bands. In addition, only a relative handful have been identified, $\approx$ 200 \citep{Belczynski00} out of an estimated 1,200-15,000 symbiotic stars with white dwarf accretors only in our Galaxy \citep{Lu06}.  SS73 17 appears to be one of an even smaller category of ``hard X-ray emitting symbiotics'', as noted above. The most significant question is why  such a small number of systems ({\it e.g.}\ SS73 17, CH Cyg, T CrB, and RT Cru) emits in hard X-rays \citep{Kennea09} while most symbiotics are faint, soft X-ray emitters \citep{Murset97}.  \citet{Kennea09} hypothesized that, unlike typical symbiotics, these systems contain particularly high-mass white dwarfs, making them potential Type Ia progenitors. 

The most likely source of the hard X-ray emission from these symbiotic stars is the same as in CVs -- the boundary layer between the accretion disk and the white dwarf \citep[\protect{\it e.g.}][]{Luna07}.  In the case of SS73 17, \citet{Smith08} noted it is a strong hard X-ray source heavily absorbed in the soft X-rays, which we confirm with these follow-up \chandra and \suzaku observations.  We detected the same strong Fe X-ray emission lines, but with some additional soft X-ray emission lines, and an increased soft X-ray continuum as well.  Our partial success with fits using a constant source and a variable partial absorber agrees with the picture suggested by \citet{Kennea09}, where the changing spectrum is due primarily to the absorbing material moving in and out of our line of sight.  

Curiously, despite the long-term spectral changes that primarily
affect the low ($<5$\, keV) energy spectrum and can therefore be described by
changes in the partial absorber, our timing analysis shows that the
short-term stochastic variations are larger when absorption is higher
(compare $s/s_{exp}$ for the first and second
\suzaku\ observations). This somehow could be linked to changes in the 
accretion rate, {\it i.e.}\ more mass flowing through the disk, increasing the
amount of absorption and increasing the viscosity in the disk. Our
analysis also points that, in general, the hard component (5--10 keV)
is more variable than its soft counterpart. In a cooling-flow
scenario, we would expect a more turbulent plasma as it gets
colder. This discrepancy, however, could be due to poor photon
statistics in the soft X-ray band, although an increase in the bin size
of the light curves, aiming at improving the signal-to-noise in each bin,
would hinder the search for short-term variability.

The inadequacies of the fit at low energies could be due to a number of causes.  We can suggest at least three possible sources for the soft X-ray emission. It could be due to photoionization of the red giant wind by the white dwarf, as proposed by \citet{WK06}, or could be created in a colliding wind shock between the M-type giant and the compact object. Alternatively, these soft X-rays could be due to a wider range of emission measures from the thermal plasma which are only partially absorbed by a thick layer of gas and dust.  With the available data, it is not possible to distinguish between these models.
  
The data do show that the hard X-rays from SS73~17 are largely thermal, due to the presence of a number of emission lines with clear thermal origin.  We found that both the forbidden ($z$) and resonance ($w$) \ion{Fe}{25} lines are present in the spectrum.  The HETG results show that the \ion{Fe}{25} lines are heavily blended with a G-ratio ($\equiv (x+y+z)/w = 1.2^{+1.1}_{-0.5}$) that supports a wide range of temperatures.  However, this value is much lower than would be expected from a photoionized plasma ($\approx 4$) \citep{PD00}.  In addition, the \ion{Fe}{26}/\ion{Fe}{25} ratio is in good agreement with the temperature inferred from a bremsstrahlung model fit to the continuum.   This thermal picture is further confirmed by the \ion{Si}{14} emission line and the likely detection of \ion{S}{16}, which have been observed in a similar symbiotic system (CH Cyg).  

The results from the \suzaku observations suggest that the source is moderately variable in the long term primarily due to changes in the absorption. This is not surprising for an X-ray binary.  However, the constancy of the Fe K$\alpha$\ fluorescent line suggests that the mechanism generating this line is largely independent of the absorption, challenging the model presented in \citet{Smith08}.  They suggested that about half (0.13 keV) of the Fe K$\alpha$ fluorescence was due to line of sight scattering in the absorbing medium, with 0.1 keV also coming from scattering off the surface of the white dwarf.  The new fits would predict only ~0.065 keV from the absorbing material, or 0.165 keV in total, significantly less than the required amount. 

Many questions remain about these unusual symbiotic systems.  We have been able to confirm the thermal origin of most of the hard X-rays, although some fraction of the flux in the 15-50 keV band -- the bandpass where this system was first noted by \integral and the \swift BAT -- remains unexplained. However, it is worth to note that INTEGRAL observations of RT~Cru were well fit with a non-thermal powerlaw emission with photon index $\Gamma$=2.7 \citep{chernyakova} and therefore we cannot discard the possibility that some fraction of the flux in 15--50 keV band from SS73~17 has a non-thermal origin.  However, it is still not at all clear if these systems are wind or disk accretors, and ultimately what drives the hard X-ray flux. More observations to determine the mass of the primary and/or secondary, as well as the orbital period of the system, would be extremely helpful in future modeling efforts.

\acknowledgments

We gratefully acknowledge helpful discussions with Koji Mukai, Norbert Schulz, David Huenemoerder, Richard Mushotzky, Tim Kallman, and Scott Kenyon.  Financial support for this work was made possible by Chandra grant GO8-9032X. {\it Facilities:} \facility{Suzaku}, \facility{CXO (HETG)}.

\end{document}